\documentclass[fleqn,usenatbib]{mnras}
\usepackage{graphicx}
\usepackage{txfonts}

\usepackage{url}
\usepackage{graphicx}
\usepackage[hyphenbreaks]{breakurl}
\usepackage{float}
\usepackage{subfig}
\usepackage{multirow}

\title[TESS observations of HMXBs]{A {\sl TESS} search for donor-star pulsations in High-Mass X-ray Binaries}

\author[]
{Gavin Ramsay,$^{1}$ Pasi Hakala,$^{2}$ Philip A. Charles$^{3}$\\
$^{1}$Armagh Observatory \& Planetarium, College Hill, Armagh, BT61 9DG, UK\\
$^{2}$Finnish Centre for Astronomy with ESO (FINCA), Quantum, FI-20014 University of Turku, Finland\\
$^{3}$School of Physics and Astronomy, University of Southampton, Highfield, Southampton SO17 1BJ, UK\\}

\date{Accepted 2022 August 3. Received 2022 August 3; in original form 2022 June 29}

\begin{document}
\outer\def\gtae {$\buildrel {\lower3pt\hbox{$>$}} \over 
{\lower2pt\hbox{$\sim$}} $}
\outer\def\ltae {$\buildrel {\lower3pt\hbox{$<$}} \over 
{\lower2pt\hbox{$\sim$}} $}
\newcommand{\Msun}{$M_{\odot}$}
\newcommand{\lsun}{$L_{\odot}$}
\newcommand{\Rsun}{$R_{\odot}$}
\newcommand{\solar}{${\odot}$}
\newcommand{\kep}{\sl Kepler}
\newcommand{\ktwo}{\sl K2}
\newcommand{\tess}{\sl TESS}
\newcommand{\swift}{\it Swift}
\newcommand{\Porb}{P_{\rm orb}}
\newcommand{\nuorb}{\nu_{\rm orb}}
\newcommand{\eplus}{\epsilon_+}
\newcommand{\eminus}{\epsilon_-}
\newcommand{\cd}{{\rm\ c\ d^{-1}}}
\newcommand{\MdotL}{\dot M_{\rm L1}}
\newcommand{\Mdot}{$\dot M$}
\newcommand{\Mdotsolar}{\dot{M_{\odot}} yr$^{-1}$}
\newcommand{\Ldisk}{L_{\rm disk}}
\newcommand{\src}{KIC 9202990}
\newcommand{\ergscm} {erg s$^{-1}$ cm$^{-2}$}
\newcommand{\rchi}{$\chi^{2}_{\nu}$}
\newcommand{\chisq}{$\chi^{2}$}
\newcommand{\pcmsq} {cm$^{-2}$}

\providecommand{\lum}{\ensuremath{{\cal L}}}
\providecommand{\mg}{\ensuremath{M_{\rm G}}}
\providecommand{\bcg}{\ensuremath{BC_{\rm G}}}
\providecommand{\mbolsun}{\ensuremath{M_{{\rm bol}{\odot}}}}
\providecommand{\teff}{\ensuremath{T_{\rm eff}}}

\maketitle
\begin{abstract}
Ground-based optical photometry of the counterparts of High-Mass
X-ray Binaries (HMXBs) has revealed the presence of periodic
modulations on timescales of $\sim$0.3-0.5 d.  More recent space-based
observations ({\sl Corot} and {\tess}) of OB and Be stars have shown
that pulsations caused by $p$ and $g$ modes are common in early type
stars. We have therefore undertaken a systematic search for
variability in the optical counterparts of 23 HMXBs (mostly neutron
star systems, but including one black hole, Cyg\,X-1) using {\tess}
data primarily in 2 min cadence mode.  After removing the orbital
period modulation in four systems, we find that all 23 sources show
evidence for quasi-periodic variability on periods shorter than
$\sim$1 d. We compare their power spectra with those from
observations of other OB and Be type stars. In two systems, V725 Tau
and HD 249179 (which may not be a HMXB), we find evidence for an
outburst, the former being simultanous with an X-ray
flare. We search for changes in the power spectra over the outburst
duration, and compare them with outbursts seen in other Be systems.
\end{abstract}


\section{Introduction}

High-Mass X-ray Binaries (HMXBs) are intense X-ray sources consisting
of a compact star, either a neutron star (NS) or a black hole (BH),
and a giant or supergiant companion star which is close to filling its
Roche Lobe. Accretion onto the primary takes place in the form of an
accretion disc and/or via a stellar wind from the hot companion. They
have orbital periods ranging from days to months and many have a high
binary eccentricity which can cause short duration X-ray outbursts on
the orbital cycle as the companion star passes through periastron
(Type I outbursts). Longer duration outbursts lasting several orbital
cycles (Type II outbursts) are seen less frequently (see
\citet{OkazakiNegueruela2001} for an overview of the outburst
models). For a detailed review of HMXBs see
\citet{LewinvanderKlis2006}.

The optical light curves of HMXBs over an interval of days and weeks
can be complex, and is a combination of the ellipsoidal modulation of
the secondary star with the effect of a tilted and precessing
accretion disc (e.g. \citet{GerendBoyton1976}). On shorter timescales,
observations by \citet{GutierrezSoto2011} of two HMXBs revealed short
period modulations which were identified as being due to 
  nonradial pulsations from the companion star in V635 Cas (0.30 d)
and GSC 03588-00834 (0.45 d). Observations using wide-field imaging
surveys, such as OGLE IV, have allowed the discovery of short period
photometric modulations in other HMXBs,
e.g. \citet{Schmidtke2014,Schmidtke2016,Schmidtke2019}. If the nature
of these periodic signals can be effectively modelled they can give
insight into the internal structure of the companion star in HMXBs.

The discovery of pulsations from the secondary star in some HMXBs is
not especially surprising since they are OB stars with spectral types
ranging from supergiants (e.g. HDE\,226868 the optical companion
to Cyg\,X-1) to dwarfs (e.g. V490\,Cep), which have previously shown
pulsations. The slowly pulsating B-type stars (SPBs) display
high-order $g$-modes (with periods $\geq$6\,hr) whilst the
$\beta$\,Cep stars stars can show lower-order $p$ modes (periods
$\sim$1--8\,hr) and $g$-modes (see \citet{DeCat2010} for a review of
observations made using {\sl Corot}). More recently,
\citet{LabadieBartz2022} made an analysis of 432 classical Be stars
observed by {\tess} during its first year of operation. Almost all of
these stars showed significant variability and their power spectra
could be classified, with 85 percent showing closely spaced
frequencies in their power spectra.

The launch of satellites such as {\sl Corot} and {\tess} have provided
a golden opportunity to search for pulsations in stars of all types,
including X-ray binaries. Although none were known in the original 115
square degree {\sl Kepler} field \citep{Borucki2010}, the prototypical
LMXB Sco\,X-1 was observed when {\kep} was repurposed as {\ktwo} with
fields along the ecliptic being observed for 2 month blocks (see
\citet{Scaringi2015,Hakala2015,Hynes2016}).  {\tess} was launched in
April 2018 and although it does not go as deep as {\kep}, it does
provide high signal-to-noise photometry of sources down to
$V\sim13-14$ (see \citet{Ricker2015} for more details). In its inital
2 year mission it observed $\sim$3/4 of the whole sky, with a gap
along the ecliptic plane and an additional section of the
northern hemisphere. Given that some HMXBs are optically bright, this
provides an opportunity to search for short period pulsations, of the
type identified by \citet{GutierrezSoto2011} and
\citet{Schmidtke2014,Schmidtke2016,Schmidtke2019}.

In this paper we present observations of the optical light curve of 23
HMXBs and identify those which show evidence for short periodic,
likely pulsation, variations. We have not attempted to do a full
frequency analysis of the light curves, but rather determine how
common pulsations are in the donor-stars of HMXBs. Further, we
identify two systems which show an optical outburst, one of which is
simultaneous with a low energy X-ray outburst. We compare these
findings with observations of other Be stars.

\section{The HMXB sample}
\label{sample}

As we are searching for periodic variability on a timescale \ltae1 d,
we restrict our sample to include HMXBs which have {\tess} 2 min
cadence data or a calibrated light curve made using full-frame image
data (see \S \ref{tess} for further details).  We cross-matched all
stars observed in Cycles 1--4\footnote{The sources in this paper were
on the 2 min cadence list thanks to their inclusion on the following
Guest Investigator programs: G011060/PI Paunzen; G011155/PI Huber;
G011204/PI Pepper; G011224/PI Redfield; G011268/PI Scaringi;
G011281/PI Silvotti; G022020/PI Dorn-Wallenstein; G022062/PI Prsa;
G022071/PI Scaringi; G022172/PI Labadie-Bartz; G022184/PI Coley;
G03156/PI Pope; G03186/PI Labadie-Bartz; G03221/PI Barlow; G04067/PI
Wisniewski; G04074/PI Bowman; G04103/PI Huber.}  with the HMXB
catalogue of \citet{Liu2006}, finding 23 sources which are detailed in
Table \ref{targetlist}. Four of our sample have supergiant donor
  stars, while more than half the sample have Be type
  donors. Although HD\,49798 is classed as HMXB in \citet{Liu2006}
this appears to be a sdO/WD binary and have therefore not included it
in this study. HD 141926 is included in the catalogue of
  candidate Herbig Ae/Be stars \citep{Vieira2003} and hence its
  evolutionary state will differ from other stars in this sample. In
\S \ref{subhd249179} we note that HD\,249179 might not be an HMXB:
however, for reasons outlined there we decided to retain this source
as part of our study.

% mid 13/24 (54%): isolated: 9/24 (38%): mid/low 2/24 8%
\begin{table*}
\caption{Properties of HMXBs observed by {\tess} in 2 min cadence
  mode.}
\label{targetlist}
\resizebox{\textwidth}{!}{
\begin{tabular}{llrrrrrrrcrl}
Name & Other Name & TIC & $T_{mag}^1$ & Sectors  & Spectral&   $P_{orb}$ & Distance$^2$ & $MG_{o}^3$ & $(BP-RP)_{o}^3$ & Primary & Power Spectra\\
     &            &     &          &          & Type     & (d)      &  (kpc)    &          &            & & Type\\
\hline
GP Vel   & Vela X-1  & 191450569 & 6.3 & 8,9 & B0.5 Ib  &  8.96  & 2.0$^{+0.1}_{-0.1}$ & -5.8 & 0.2 & NS & Mid \\
Hen 3-640 & 1A 1118-615& 468095832 & 10.7 & 10,37,38 & O9.5Ve    & 24.0 & 3.0$^{+0.2}_{-0.1}$ & -2.0 & 0.9 & NS & Mid/Low\\
V801 Cen & 2S 1145-619 & 324268119 & 8.7 & 10,37,38 & B0.2IIIe  & 187.5  & 2.1$^{+0.1}_{-0.1}$ & -4.0 & 0.0 & NS & Mid\\
BZ Cru & 1H 1249-637 & 433936219 & 5.2 & 11,37,38 & B0.5IIIe &  - & 0.44$^{+0.03}_{-0.02}$ & -3.4 & 0.4 & NS & Mid\\
$\mu^{2}$ Cru  & 1H 1255-567&  261862960 & 5.3 & 11,37,38 & B5Ve    &  - & 0.121$^{+0.003}_{-0.003}$ & -0.4 & -0.2 & & Mid \\
HD 141926 & 1H 1555-552  & 84513533  & 8.0 & 12,39 & B2IIIn & - & 1.4$^{+0.1}_{-0.1}$ & -2.7 & 0.5 & & Isolated \\
V884 Sco & 4U 1700-37 &  347468930 & 6.1 & 12 & O6.5Iaf & 3.41 & 1.6$^{+0.2}_{-0.1}$ & -5.3 & 0.1 &  & Mid\\
HDE\,226868  & Cyg X-1 &  102604645 & 7.9 & 14 & O9.7 Iab &  5.6  & 2.2$^{+0.2}_{-0.1}$ & -4.3 & 0.7 & BH & Mid \\
GSC 03588-00834 & SAX J2103.5+4545  & 273066197 & 13.0 &  16 & B0Ve     &  12.68 & 7.4$^{+1.5}_{-0.9}$ & -2.3 & 0.7 & NS & Mid\\
V490 Cep & 1H 2138+579 &  341320747 & 13.0 & 16,17  & B1-B2Ve  & &  9.0$^{+2.1}_{-1.2}$ & -2.8 & 0.6 & NS & Mid \\
BD +53 2790 & 4U 2206+543 &  328546890 & 9.6 & 16,17 & O9.5Ve   &  9.57 & 3.3$^{+0.2}_{-0.2}$ & -4.1 & -0.2 & NS & Mid/Low\\
V662 Cas & 2S 0114+650   & 54469882  & 9.8 & 18,24,25 & B0.5Ib &  11.6 & 5.1$^{+0.5}_{-0.4}$ & -4.5 & 0.8 & NS & Mid \\
V635 Cas & 4U 0115+634  & 54527515  & 13.3 & 18,24,25 & B0.2Ve  &  24.3 & 7.0$^{+1.8}_{-1.0}$ & -1.7 & 1.3 & NS & Isolated\\
BQ Cam & EXO 0331+530 &  354185144 & 13.0 & 18,19 & O8.5Ve  &  34.25 & 7.0$^{+2.2}_{-1.1}$ & -1.8 & 1.5 & NS & Mid \\
X Per & 4U 0352+309    & 94471007  & 6.0 & 18,43,44 & B0Ve    &  250.3 & 0.61$^{+0.03}_{-0.02}$ & -3.2 & 0.3 & NS & Mid \\
CI Cam & XTE J0421+560  & 418090700 & 9.9 & 19 & sgB[e]  & 19.41 & 4.7$^{+0.7}_{-0.5}$ & -4.0 & 0.9 & & Isolated\\
LS V +44 17  & RX J0440.9+4431 &  410336237 & 10.0 & 19  & B0.2Ve  & 150.0 & 2.6$^{+0.2}_{-0.1}$ & -2.8 & 0.5 & NS & Mid\\
 V420 Aur & EXO 051910+3737.7 & 143681075 & 7.0 &  19  & B0 IVpe & & 1.4$^{+0.1}_{-0.1}$ & -4.2 & 0.1 & &  Isolated\\
 HD 109857$^4$ & 1H 1253-761 & 360632151 & 6.5 & 11,12,38,39 & B8V & & 0.21$^{0.02}_{-0.02}$ & -0.3 & 0.1 & & Isolated \\
V725 Tau & 1A 0535+262 & 75078662 & 8.2 & 43-45 & O9/B0III/Ve & 111.0 & 1.91$^{0.16}_{-0.12}$ & -3.7 & 0.5 & NS & Isolated\\
IGR J06074+2205 & & 45116246 & 11.8 & 43,44 & B0.5Ve & & 6.9$^{+1.9}_{-1.0}$ & -3.8 & -0.1 & & Isolated\\
BD+47 3129 & RX J2030.5+4751 &  187940144 & 8.7 & 41 & B0.5V-IIIe & & 2.39$^{+0.17}_{-0.13}$ & -3.9 & 0.2 & & Isolated\\
HD 249179 & 4U 0548+29 & 78499882 & 9.2 & 43-45 & B5ne & & 1.68$^{+0.16}_{-0.12}$ & -2.0 & -0.3 & & Mid\\
\hline
\end{tabular}}
{\footnotesize
$^1$ Magnitude in the {\tess} pass-band;
$^2$ From {\it Gaia} EDR3;
$^3$ De-reddened;
$^4$ Data taken in 30 min cadence.  
}
\end{table*}

For further insight as to the nature of the donor stars in these 23
HMXBs, we use the {\sl Gaia} EDR3 parallaxes \citep{Gaia2021} to infer
their distances\footnote{ following the guidelines of
\citet{BailerJones2015,Astra2016} and \citet{GaiaLuri2018}, which is
based on a Bayesian approach.}. In practise we use a routine in the
{\tt STILTS} package \citep{Taylor2006} and use a scale length L=1.35
kpc, which is appropriate for stellar populations in the Milky Way in
general.  From these distances we determine their absolute magnitudes
in the Gaia $G$ band (a very broad optical filter), $M_{G}$, using the
mean Gaia $G$ magnitude.  The other key observables are the blue
($BP$) and red ($RP$) filtered magnitudes, which are derived from the
Gaia Prism data. We then deredden the $(BP-RP)$, $M_{G}$ values using
the 3D-dust maps derived from Pan-STARRS1 data \citep{Green2019} and
the relationship between $E(B-V)$, $E(BP-RP)$, $A_{G}$ and
$E(BP-RP)$ outlined in \citet{Andrae2018}. For those stars just off
the edge of the Pan-STARRS1 field of view (stars with
$\delta<-30^{\circ}$) we take the nearest reddening distance
relationship.  We include the distances to our targets and $MG_{o}$,
$(BP-RP)_{o}$ in Table \ref{targetlist}.

\begin{figure}
    \centering
    \includegraphics[width = 0.45\textwidth]{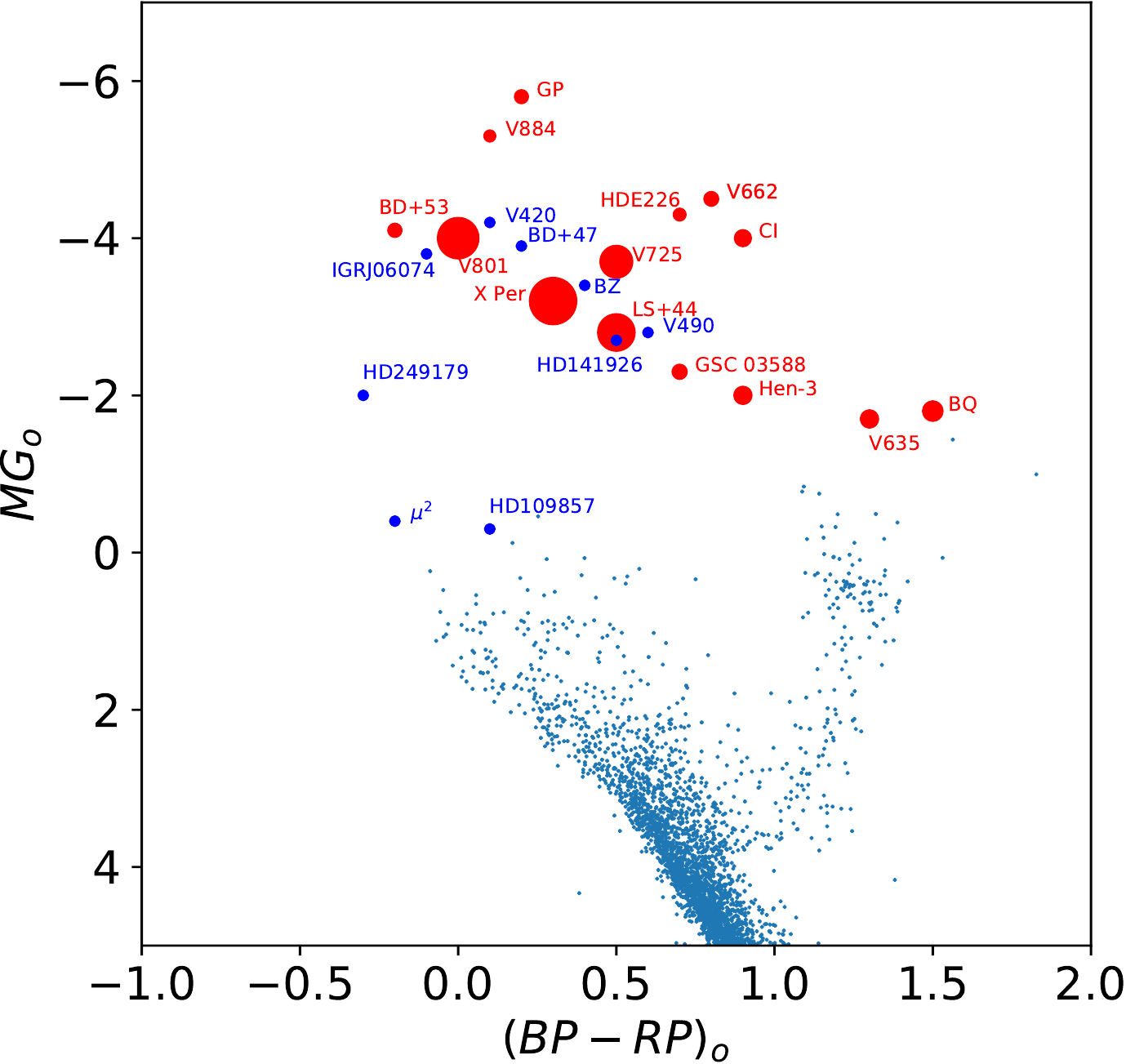}
    \caption{The Gaia HRD ($(BP-RP)_{o}$, $MG_{o}$) where the blue
      dots represent stars within 50 pc of the Sun and we assume they
      have negligible reddening. The HXMBs in our sample are shown as
      circles with a shortening of their name, and are dereddened as
      outlined in the main text. Blue circles indicate those with
      unknown $P_{\rm orb}$, whilst the size of the red circles are
      proportional to $P_{\rm orb}$.}
    \label{gaia-hrd}
\end{figure}

The nearest HMXB in our sample is $\mu^{2}$\,Cru which is only 120\,pc
distant (we note that although $\mu^{2}$\,Cru is bright,
  $T_{mag}$=5.3, there is another star 35.3$^{''}$ distant which is
  $T_{mag}$=4.2, so some contamination in the light curve will be
  present). In contrast, V490\,Cep (1H\,2138+579) is $\sim$9\,kpc
distant, although with a large uncertainty. We show the dereddened
$MG_{o}$ and $(BP-RP)_{o}$ values for our sample in Figure
\ref{gaia-hrd}, where the size of the symbols reflect the binary
orbital period (if known). To give context, we also show the apparent
$M_{G}$ and $(BP-RP)$ values for stars within 50\,pc and assume they
show negligible reddening. What is immediately striking is the spread
in position of our sample HMXBs in the Gaia HRD. Cyg\,X-1, the one
system which we know has a BH primary is near the upper part of the
distribution. Those sources close to Cyg\,X-1 are V662\,Cas (a NS
primary with a very slow rotation period of 2.78\,hr \citep{Hall2000})
and CI\,Cam the nature of whose primary remains controversial.

\section{The TESS data}
\label{tess}

\begin{figure*}
    \centering
    \includegraphics[width = 0.95\textwidth]{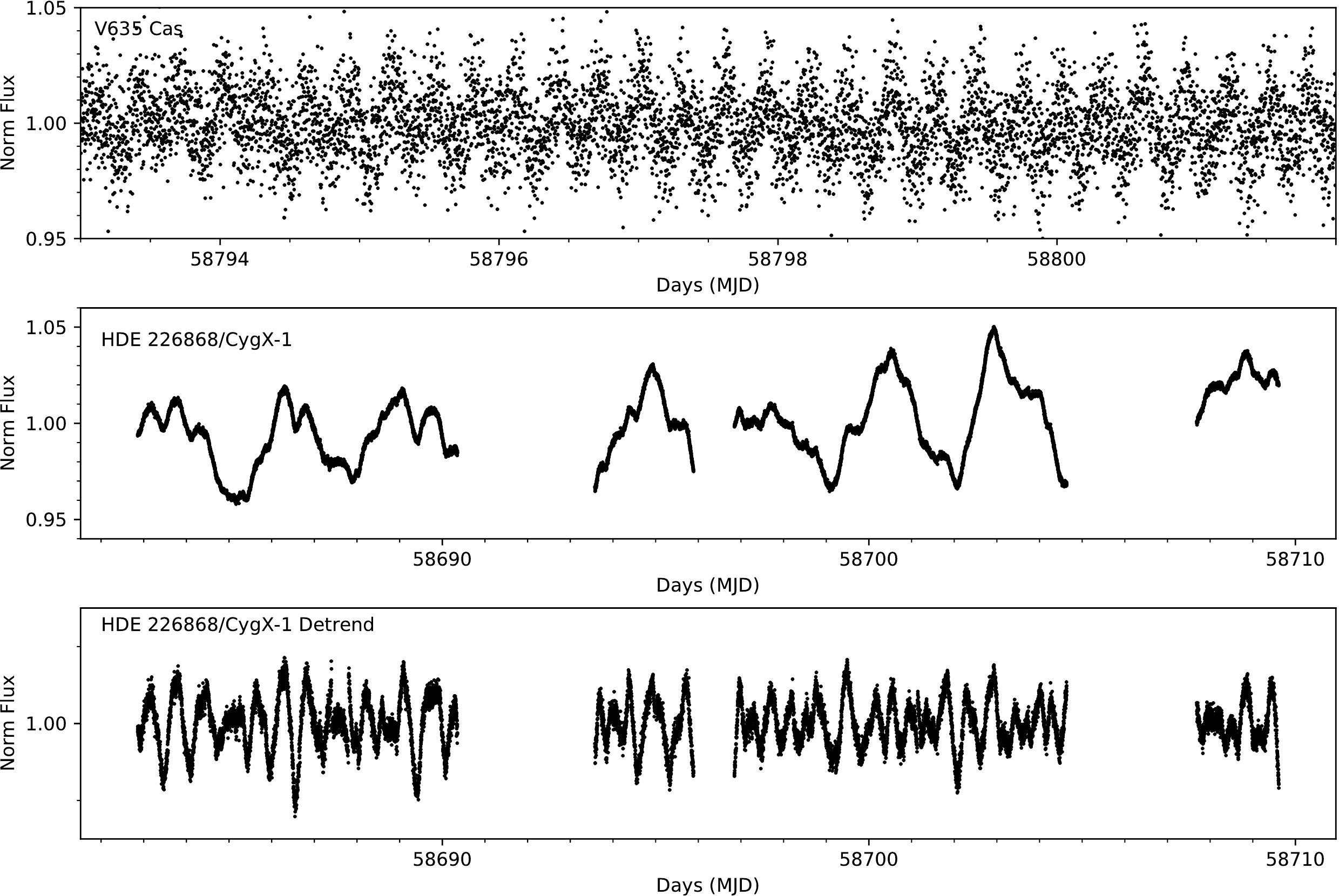}
    \caption{Upper Panel: A nine day section of {\tess} 2 min cadence
      data of V635\,Cas from sector 18. There is a clear modulation on
      a period of 0.30 d. Middle Panel: The {\tess} 2 min cadence data
      of HDE\,226868 (Cyg X-1) obtained in sector 14. A high amplitude
      modulation is seen on the orbital period (5.6 d) and half the
      orbital period. Lower Panel: The light curve of HDE\,226868
      detrended to remove the signature of the orbital period.}
    \label{v635cas-cygx1}
\end{figure*}

{\sl TESS}, launched on 18th April 2018, has four telescopes, each
with an aperture of 10.5\,cm that cover a 24$^{\circ}\times90^{\circ}$
sector of sky for $\sim$27\,d. In the prime mission, the
majority of the sky was observed apart from a strip along the ecliptic
plane. In the first year of the extended mission, the southern
ecliptic hemisphere was observed for a second time, whilst in the
second year those regions not observed in the prime mission were
covered with a second observation of northern fields.

The {\tess} camera detectors have $21^{``}\times21^{``}$ pixels, with
the PSF covering more than 2$\times$2 pixels, which can make blending
an issue especially in crowded fields. For the first two years,
$\sim$20,000 predefined targets were observed in each sector with a
2\,min cadence, with a 20\,s cadence being introduced in year 3. In
years 1 and 2, full-frame images with a cadence of 30\,min are
available, with the cadence increasing to 10\,min in year 3.  All of
our targets, with the exception of HD\,109857 (which has 30\,min
cadence in sectors 11--12 and 10\,min in sectors 38--39), were
observed in 2\,min cadence.

We downloaded the calibrated light-curves with 2\,min cadence of our
targets from the MAST data
archive\footnote{\url{https://archive.stsci.edu/tess/}}. We used the
data values for {\tt PDCSAP\_FLUX}, which are the Simple Aperture
Photometry values, {\tt SAP\_FLUX}, after the removal of systematic
trends common to all stars on that Chip. Each photometric point is
assigned a {\tt QUALITY} flag which indicates if the data has been
compromised to some degree by instrumental effects. For HD\,109857, we
obtained calibrated light-curves from full-frame images using the
TESS-SPOC pipeline \citep{Caldwell2020}, again from the MAST data
archive\footnote{\url{https://archive.stsci.edu/hlsp/tess-spoc}}. We
removed those points which did not have {\tt QUALITY=0} and normalised
each light-curve by dividing the flux of each point by the mean flux
of the star in that sector.

As an example of the light curves obtained by {\tess}, we show in
Figure \ref{v635cas-cygx1} V635\,Cas, which has a clear 
  short-period modulation with the same period (0.3 d) as found by
  \citet{GutierrezSoto2011}, and HDE\,226868 (the optical counterpart
to Cyg\,X-1).  The latter displays the signature of the 5.6\,d orbital
period through the well-known ellipsoidal modulation of the donor star
(hence its appearance as peaks every 2.8\,d).  Even though Cyg\,X-1 is
an extremely luminous ($\geq$ 10$^{37}$ erg s$^{-1}$) X-ray source,
the donor is an OB I star, so the effects of X-ray heating are
small. (As a benefit to the reader we show one sector of {\tess} data
for each of our targets in Fig. \ref{lightcurves-A} to
\ref{lightcurves-D}).

Four sources showed evidence of orbital modulations (GP\,Vel,
V884\,Sco, HDE\,226868 and V662\,Cas). We detrended these light curves
using the {\tt flatten} routine in the {\tt lightkurve} python package
\citep{lightkurve2018} to search for shorter period variability.
After some trial and error we chose a window length for the filter of
1\,d: we acknowledge that the resulting detrended light curves could
contain features which are a result of this choice of window
length. Nevertheless, we show in the lower panel of Figure
\ref{v635cas-cygx1} the detrended light curve of HDE\,226868, which
reveals significant short period variability in the light curve (we
show the normalised and detrended light curves for GP Vel, B884 Sco
and V662 Cas in Figure \ref{detrendappendix}).  In sector 44, X\,Per
shows a prominent sinusoidal variation on a timescale of $\sim$15\,d
(see Fig. \ref{lightcurves-C}) which is much shorter than its orbital
period of 250 d. We also removed the 15 d trend in X Per to search for
shorter period variability.

To search for periodic variations in the light curves, we used the
generalised Lomb Scargle periodogram
\citep[LS,][]{Zechmeister2009,Press1992}, obtaining a power spectrum
for each sector the star was observed. We show the power spectra
obtained for each sources in Figures \ref{power-A} to \ref{power-C}
with the periods of the most prominent peaks shown in Table
\ref{lsperiods} on a sector-by-sector basis.

In their study of 432 classical Be stars made using {\tess} data
\citet{LabadieBartz2022} classified them based on their power
spectra. The classifications include low frequency ($>$2\,d) signals
which dominate and are typical of $g$ mode pulsations; frequency
groups in the mid frequency range (0.16--2\,d) closely spaced groups
($p$ and $g$ modes); high frequency (0.06--0.16\,d) signals dominate
($p$ and $g$ modes); and single isolated frequencies -- sources can
have more than one classification. We made an assessment of the power
spectra shown in Figures \ref{power-A} to \ref{power-C} and classified
them based on the criteria of \citet{LabadieBartz2022}. We find that
13/23 (57 percent) have mid frequency spectra;  8/23 (35
  percent) have isolated spectra and 2/23 (9 percent) show both
mid and low frequency spectra. In comparison, \citet{LabadieBartz2022}
find that 32 percent of Be stars show isolated frequencies and 85
percent show frequency groups. {Hen 3-640 was included in both
  this study and that of \citet{LabadieBartz2022}, with us classifying
  the power spectra as mid/low and \citet{LabadieBartz2022} indicating
  it shows low-frequency trends and isolated signals (including at
  high frequencies).  Given the element of uncertainty in classifying
the power spectra of the HMXB in our sample, the light curves analysed
here appear to be broadly similar to the classical Be stars reported
in \citet{LabadieBartz2022}.

In Figure \ref{gaia-hrd-power-type} we show the same Gaia HRD as we
showed in Figure \ref{gaia-hrd} but here we have colour coded the
sources depending on their classification shown in Table
\ref{targetlist}. There appears to be no separation between sources
showing mid frequency groups, isolated signals and those showing mid
and low frequency signals.

We now comment on some specific sources.  GSC 03588-00834 (SAX
  J2103+4545) shows two eclipses in its {\tess} light curve separated
  by 16.6 d. Given this is longer than the orbital period, this
suggests that one of two relatively bright stars ($G$=13.9, $G$=14.3)
within 21 arcsec (the {\tess} pixel size) is the source of the
eclipses. The light curve also shows evidence of two dip-like features
which may be related to the orbital period. Using a Weighted Wavelet
Z-Transform \citep{Foster1996} to search for evidence of short period
variations we found evidence for a period which drifts in the three
sectors of data between 0.33--0.45 d which is similar to the period
reported by \citet{GutierrezSoto2011}. However, once we take into
account the dilution of the two spatially nearby stars this increases
to 1.2 percent. We conclude that it is likely that a quasi-periodic
signal on a timescale similar to that reported by
\citet{GutierrezSoto2011} is present at times in GSC
  03588-00834.

We noted earlier that X Per shows a modulation on a timescale of
$\sim$15 d in sector 44 whose signature we removed before obtaining a
LS power spectrum. In Figure \ref{power-B} we see that power
is concentrated at periods near 0.50-0.55 d in sectors 18 and 43, but
in sector 44 the peak in the power spectrum is shifted to an isolated
peak at $\sim$0.3 d (there is no evidence for an X-ray outburst at
this time). In Figure \ref{power-A} we show the power spectra
of Hen 3-640: in sector 10, peaks are seen between periods of 0.5-1.1
d but in sectors 37 and 38 there is a clear modulation in a period
$\sim$ 3 d. Similarly, in Figure \ref{power-A} we find that
BZ Cru shows enhanced power between 0.6-0.8 d in sectors 37 and 38
compared to sector 11. For two sources, V725 Tau and HD 249179, we
discuss their power spectra in greater detail in \S \ref{outburst}.

\begin{figure}
    \centering
    \includegraphics[width = 0.45\textwidth]{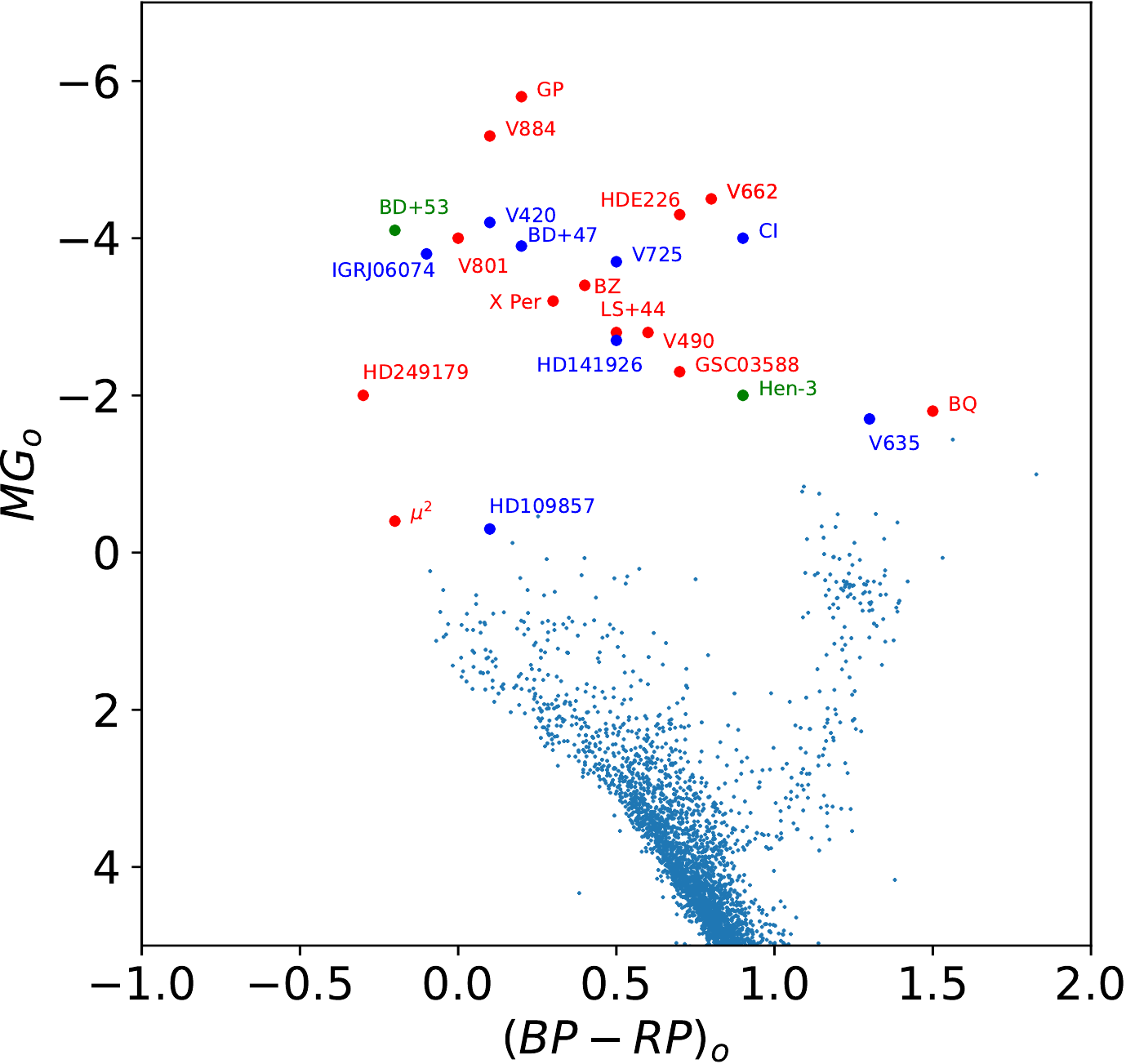}
    \caption{The Gaia HRD as shown in Figure \ref{gaia-hrd} but here
      with the colour of the symbol representing those sources with
      mid frequency signals in red; isolated signals in blue and those
      showing mid and low frequency signals in green.}
    \label{gaia-hrd-power-type}
\end{figure}

\begin{table*}
  \begin{tabular}{lr}
    Object         & Periods (d) \\
\hline
GP Vel             &  [8] 0.521 [9] 0.656 \\
Hen 3-640          &  [10] 1.066 [37] 3.06  [38] 3.05, 1.072 \\
V801 Cen           &  [10] 0.665, 0.357, 0.464 [37] 0.358, 0.463, 0.661  [38] 0.463, 0.358, 0.375 \\
BZ Cru             &  [11] 0.339, 0.104, 0.969 [37] 0.725, 0.637, 0.694 [38] 0.636, 0.339, 0.725 \\
$\mu^{2}$ Cru      & [11] 1.614, 0.935, 0.989, 0.336 [37] 0.918, 1.005, 0.337, 1.066 [38] 1.598, 0.930, 0.337, 0.893 \\
HD 141926           &  [12] 0.904, 0.953, 0.419 [39] 0.904, 0.407, 0.415 \\
V884 Sco           & [12] 0.729, 0.763, 0.614 \\
HDE 226868          & [14]  0.661, 0.597, 0.578 \\
GSC 03588-00834     & [16] 0.448, 0.102 \\
V490 Cep           & [16] 1.970, 0.538 [17] 1.963, 1.772 \\
BD +53 2790        & [16] 1.120, 0.981, 0.647 [17] 0.998, 0.834, 0.791 \\
V662 Cas           & [18] 0.497, 0.352, 0.572 [24] 0.687, 0.741, 0.791  [25] 0.594, 0.613, 0.578 \\ 
V635 Cas           & [18] 0.3003 [24] 0.3003 [25] 0.3010 \\
BQ Cam             & [18] 0.419, 0.755 [19] 0.420, 0.756 \\
X Per              & [18] 0.574, 0.277  [43] 0.575, 0.277 [44] 0.277, 0.529 \\
CI Cam             & [19] 0.406 \\
LS V +44 17        & [19] 0.385, 0.460\\
V420 Aur           & [19] 0.672, 1.446  \\
HD 109857           & [11] 0.568, 0.284 [12] 0.569, 0.285  [38] 0.571, 0.454, 0.285 [39] 0.570, 0.284, 0.453 \\
V725 Tau           & [43] 0.468 [44] 0.469 [45] 0.468 \\
IGR J06074+2205         & [43] 0.434 [44] 0.435 \\
BD +47 3129          & [41] 1.225 \\
HD 249179           & [43] 0.283  [44] 0.490, 0.283 [45] 0.282  \\
\hline
  \end{tabular}
\caption{The principal periods determined for our targets on a
  sector-by-sector basis (indicated by the square brackets) using the
  LS periodogram.}
\label{lsperiods}
\end{table*}

\begin{figure*}
%\vspace{0.5cm}
  \centering
    \includegraphics[width = 0.9\textwidth]{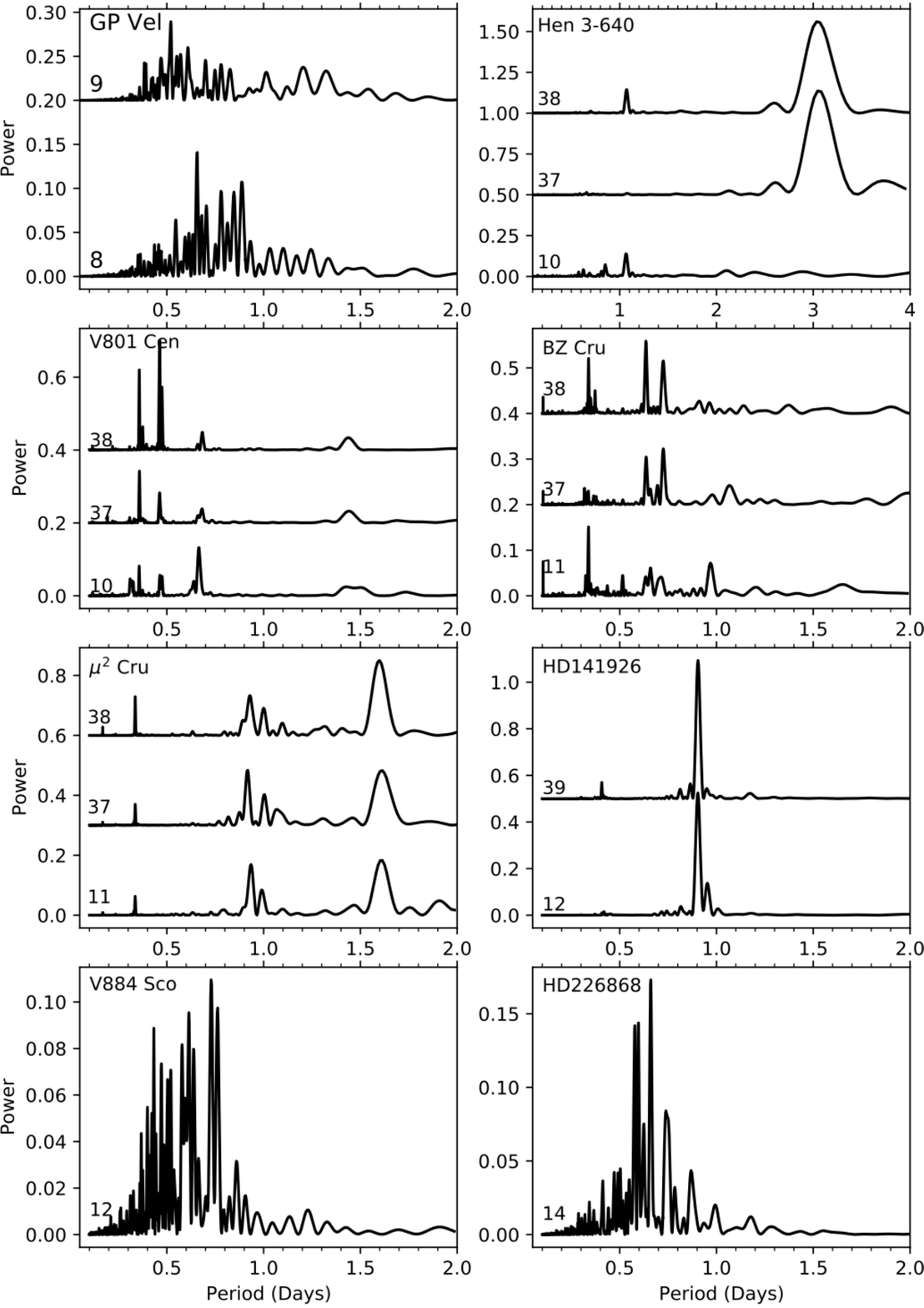}
    \caption{The LS power spectra of the {\tess} light curves of our HMXB
      sample, each labelled with the sector where the data originated.}
    \label{power-A}
\end{figure*}

\begin{figure*}
%\vspace{0.5cm}
  \centering
    \includegraphics[width = 0.9\textwidth]{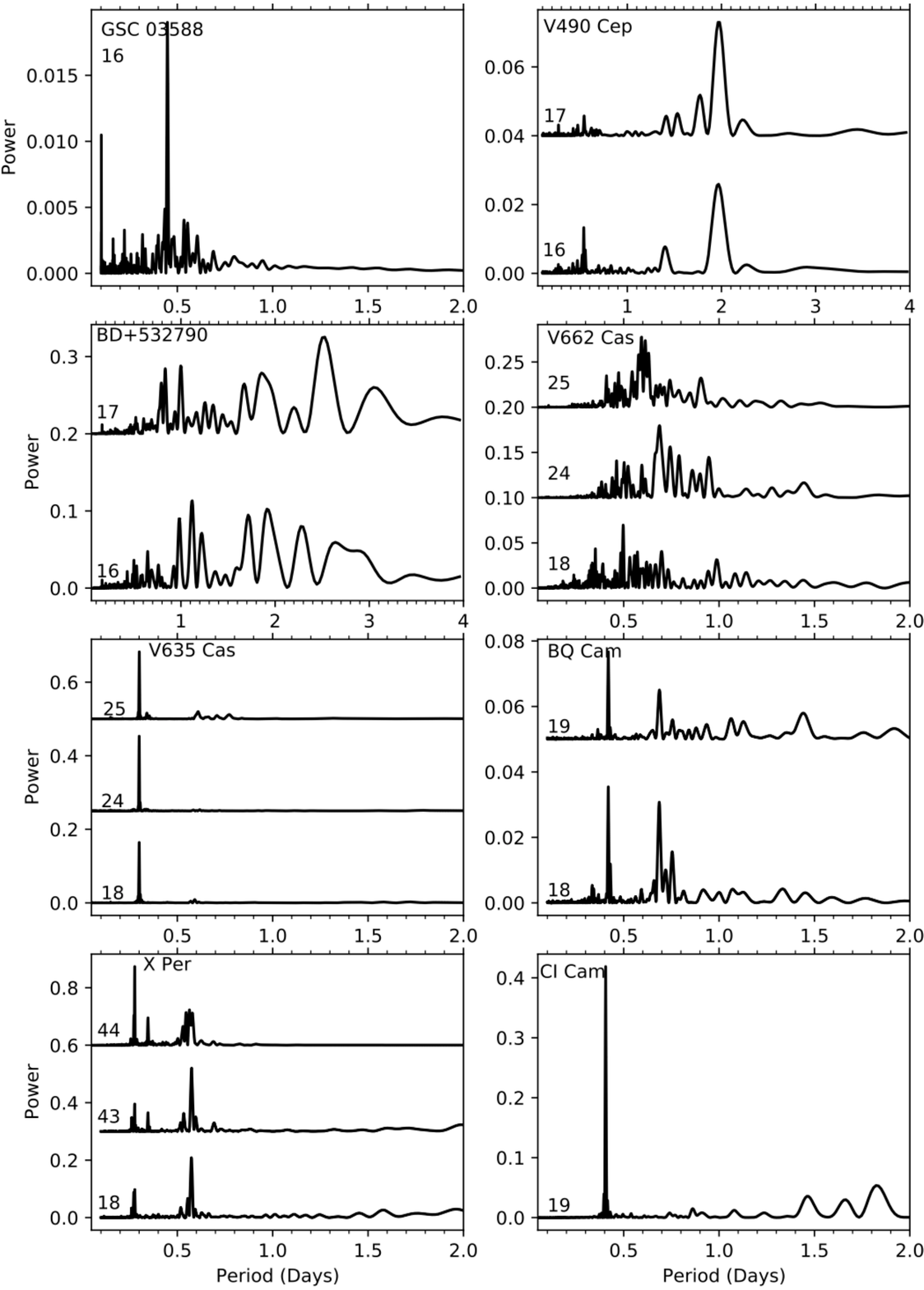}
    \caption{Figure \ref{power-A} continued.}
    \label{power-B}
\end{figure*}

\begin{figure*}
%\vspace{0.5cm}
  \centering
    \includegraphics[width = 0.9\textwidth]{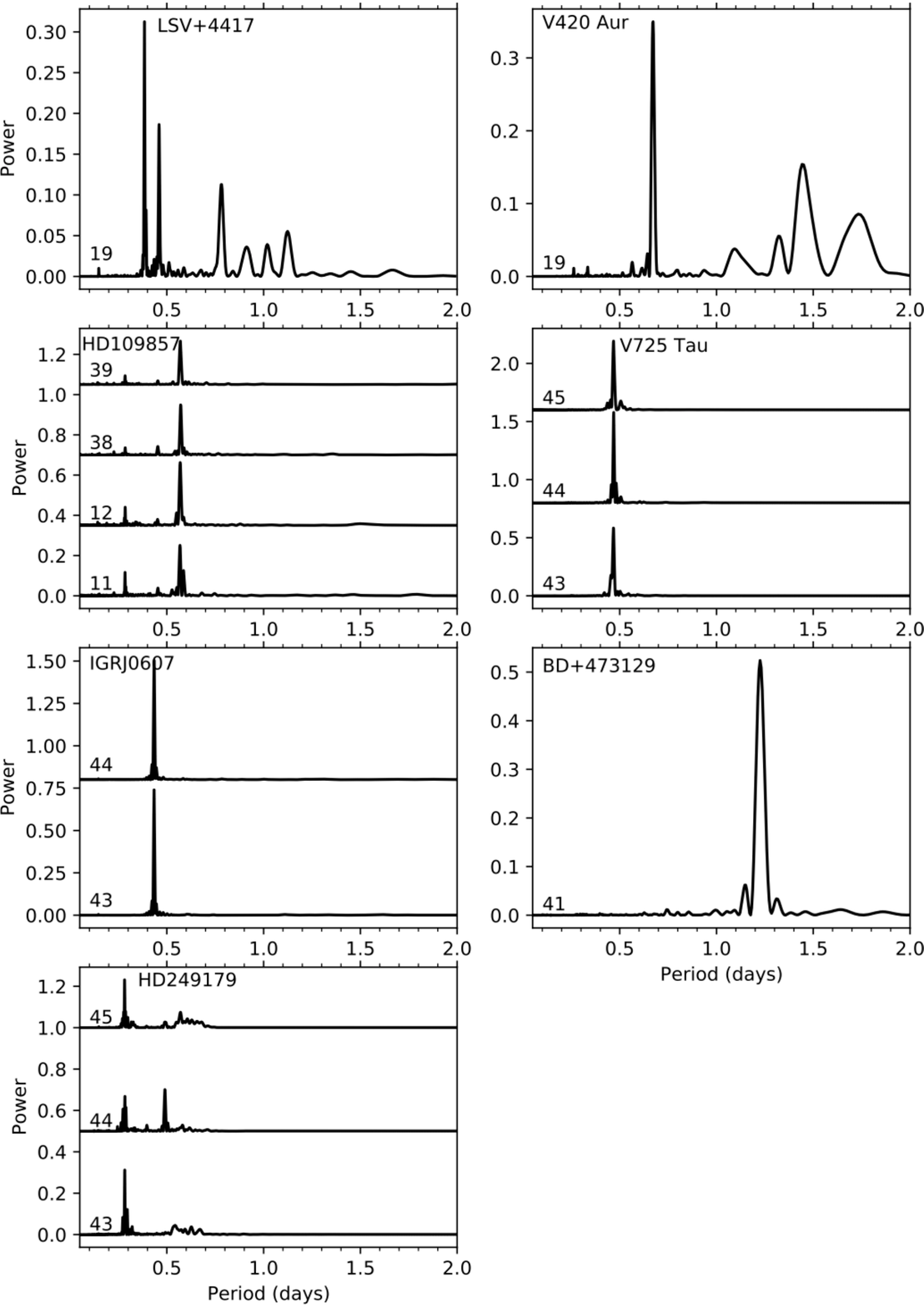}
    \caption{Figure \ref{power-B} continued.}
    \label{power-C}
\end{figure*}

\section{Outbursts}
\label{outburst}

{\tess} has been used to observe supernovae prior to their detection
in ground-based transient surveys (e.g. \citet{Fausnaugh2021}) and
also serendipitous outbursts of previously known Cataclysmic Variables
(e.g. \citet{Court2019}). Moreover, outbursts have been seen in many
classical Be stars. We therefore manually searched the {\tess} light
curves in our sample for any evidence of outbursts. We found outbursts
in two of them, V725\,Tau and HD\,249179.

\subsection{V725\,Tau}

V725\,Tau (1A\,0535+262/HD\,245770) is a prototypical transient X-ray
pulsar which has been observed to show many X-ray outbursts, including
a recent giant outburst (peaking at 1.2$\times10^{38}$ erg s$^{-1}$)
in Nov 2020 (e.g. \citet{Kong2021}). Since then, observations made
using MAXI showed a weaker outburst around 2021 June 20 (the field
was not being observed by {\tess}) and an even weaker X-ray outburst a
few months later, starting on Oct 13 (MJD=59500).  In Figure
\ref{outburst} (left) we show the simultaneous {\tess} and X-ray
(MAXI) light curves of V725\,Tau during this much weaker outburst. The
increase in optical flux suggests an outburst amplitude of $\sim$0.2
mag (there are no significant issues of dilution from nearby bright
stars).

Figure \ref{outburst} indicates there was a clear periodic ($\sim$0.5
d) signal in all three sectors (43--45) of {\tess} data. To
investigate this further, we detrended each of these sectors so as to
remove the large outburst variation, and show the resulting light
curve in Figure \ref{V725-detrend}: it appears like a typical
multi-periodic pulsation where modes come and go on a quasi-period of
$\sim$6.4\,d, causing a change in the amplitude of the main
pulsation. We applied a shifted time window version of the epoch
folding periodogram \citep{Davies1990} to search for the evolution of
the pulsations over the TESS observations. Firstly, long term
variations were removed, after which we moved a 2\,d wide time window
in steps of 0.05\,d through the observations and carried out the epoch
folding analysis with 30 phase bins for each step. We show the
resulting trailed periodogram in the upper left hand panel of Figure
\ref{outburst}. There is no apparent change in the amplitude of the
$\sim$0.5 d period during or immediately after the optical
outburst. However, it is clear that we detect at least three epochs of
increased pulsation amplitude, separated by $\sim$ 30\,d. An Analysis
of Variance (AoV) periodogram \citep{Schwarzenberg1996} using the full
detrended light curve finds a peak in the power spectrum at 13.6\,d:
this is half of the reported period of $\sim$28\,d determined using
radial velocities \citep{Hutchings1978}. Given the orbital period is
111\,d, it is unclear what the origin of the 28\,d period is due to.

\begin{figure*}
\centering
\includegraphics[width = 1.0\textwidth]{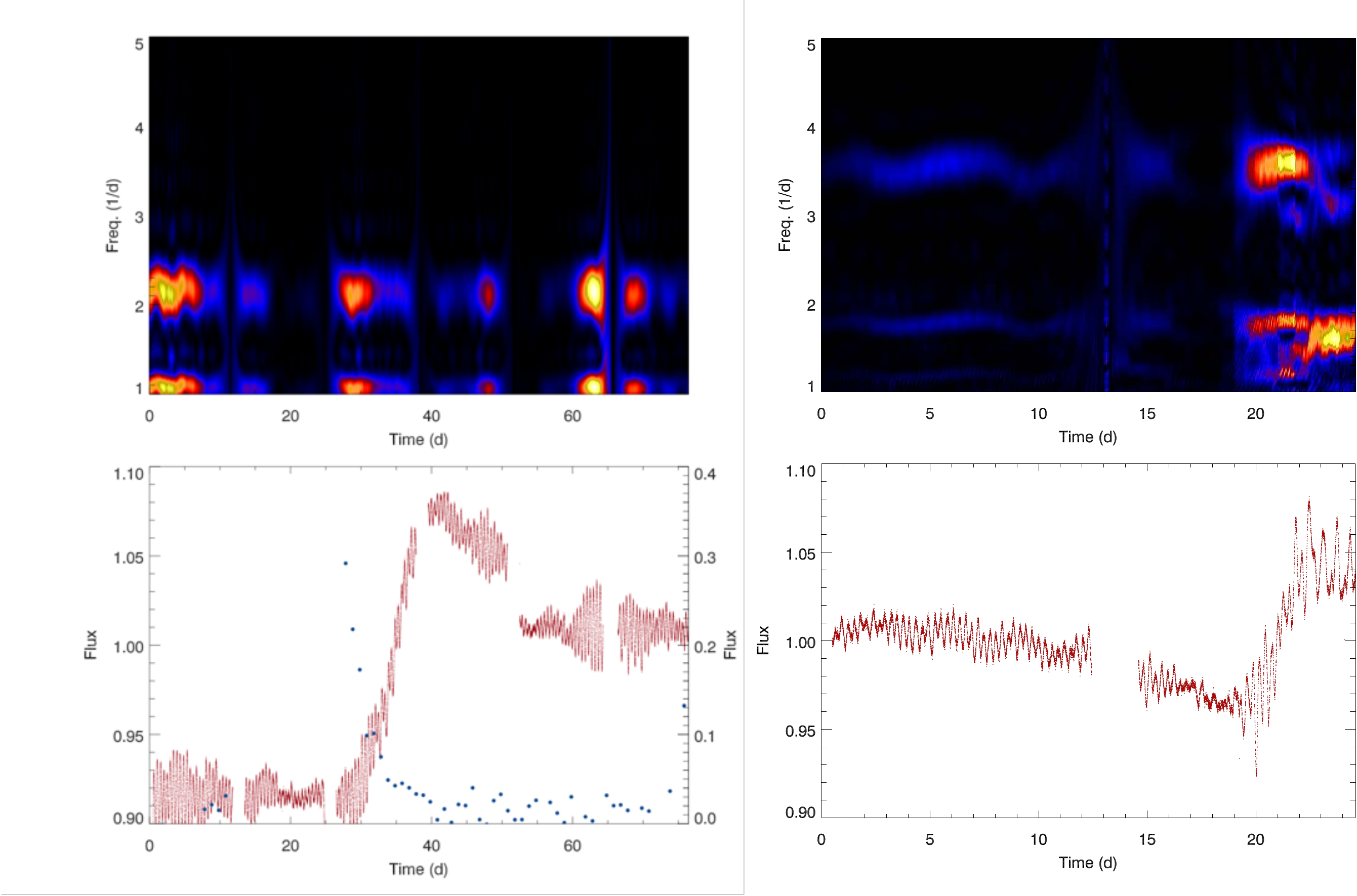}
    \caption{Lower panels: An outburst of V725\,Tau (left hand panel)
      which started on 2021 Oct 13 (day zero corresponds to
        MJD=59473.68) during {\tess} sector 44, together with the
      MAXI X-ray data (blue dots). The right hand panel shows an
      outburst of HD\,249179 during {\tess} sector 45 (day zero
        corresponds to MJD=59525.54). Upper panels: Sliding
      periodograms of the {\tess} light curves (V725 Tau, left hand
      panel, HD\,249179, right hand panel) where the colour intensity
      is scaled to reflect the amplitude of the power. }
    \label{outburst}
\end{figure*}

\subsection{HD 249179}
\label{subhd249179}

4U\,0548+29 \citep{Forman1978} was associated with the B5ne star
HD\,249179 by \citet{Wackerling1970} and classed as an HMXB in the
catalogue of \citet{Liu2006}. However, \citet{Torrejon2001} made
pointed observations of HD\,249179 using BeppoSax and found no
evidence for X-ray emission, concluding that HD\,249179 was not in
fact the optical counterpart of 4U\,0548+29.  Nevertheless, we have
retained HD\,249179 in this study as a potential HMXB, since Be stars
are known to undergo long intervals of weak or no X-ray emission, as a
result of their highly eccentric orbits \citep{OkazakiNegueruela2001}.

In Figure \ref{outburst} (right hand panels) we show the light curve
of HD\,249179 during sector 45: in the second half, after a slight
decline in flux, there is a rapid increase of $\sim$0.1 mag (this
increases to $\sim$0.14 mag once dilution from nearby bright stars is
taken into account).  We believe this is the first recorded optical
outburst from HD\,249179.

The detrended light curve of HD\,249179 (Figure
\ref{HD249179-detrend}) shows a quasi-periodic signal on a period of
$\sim0.28$\,d. This light curve also highlights how the amplitude of
the short period modulation increases by a factor of $\sim$3 at the
same time as the start of the optical outburst. This is also
demonstrated in the sliding epoch folding periodogram shown in the
upper right hand panel of Figure \ref{outburst}. In addition there is
some evidence that after the peak of the outburst has been reached,
most of the power is transferred from the initial 0.28\,d period to
twice that (i.e. from 3.6 d$^{-1}$ to 1.7 d$^{-1}$
in frequency). We speculate that material has been ejected from this
Be star to its disc.

\section{Discussion}

We now discuss pulsations from isolated early type stars and previous
observations of outbursts from Be type stars.

\subsection{Pulsations from early type stars}

\citet{Bowman2020} gives an overview of observations of High-Mass
stars including $\beta$\,Cep and SPB stars.  {\tess} observations of
nearly 100 OB stars, coupled with high resolution optical
spectroscopy, were presented by \citet{Burssens2020}, who found that
many of them showed pulsations and concluded that if the modes of
pulsation could be determined then asteroseismic modelling would be
possible. They identified a group of pulsators which show periods of a
few days, such as SPBs which were likely due to coherent $g$ mode
pulsations. Furthermore $\beta$\,Cep stars show periods of a few
tenths of a day, which are likely due to coherent $p$ mode
pulsations. In addition, there are a group of stars which appear to be
hybrid pulsators, showing both $g$ and $p$ mode
pulsations. \citet{Sharma2022} presented {\tess} observations of 119 B
type stars which were members of the Sco-Cen association and found
pulsations in 2/3 of the stars they sampled. Although they initially
applied a cut-off at 0.4\,d where stars were likely to be SPB if the
main periods were $<$0.4\,d and $\beta$\,Cep if $>$0.4\,d, they
conclude that it was not always possible to separate these stars
purely on this criterion. Indeed there were stars which appeared to be
hybrids showing power in both frequency ranges.

However, based on these criteria, we find that 42\% of our sample were
hybrids, 33\% were SPBs and 25\% were $\beta$\,Cep stars. This compares
with 23\% hybrids, 28\% SPBs and 40\% $\beta$\,Cep stars as we have
estimated from Table 1 of \citet{Sharma2022}. Given the indirect
nature of the comparison, and that the \citet{Sharma2022} study covers
all B spectral sub-types, whilst our sample is more concentrated on
earlier sub-types or even giant/sub-giants, there is not a large
difference in the fractions of types in these two studies.

\subsection{Outbursts}

Optical outbursts have been seen from many Be stars with timescales
ranging from days, weeks to years (e.g. \citet{LabadieBartz2018}). The
outburst causes material to be expelled from the star and forms (or
maintains) a {\sl excretion} disc. Much of this material is eventually
returns onto the star (see \citet{Rivinius2013} for details).

Using {\tess} observations we have detected optical outbursts from
V725\,Tau and HD\,249179.  V725\,Tau is a {\em bona fide} HMXB with a
neutron star as it primary. In contrast, there is some doubt that
HD\,249179 is an HMXB, but rather an isolated Be star
\citep{Torrejon2001}. The outburst from V725\,Tau showed no change in
the period or amplitude of the short period modulation which is likely
due to $p$-mode oscillations in the donor star. In contrast, we have
found evidence that the period changes slightly during the outburst of
HD 249179, with strong evidence that the amplitude increases
significantly at the time where the outburst starts.

The outburst which we detected in HD\,249179 is very similar to the
outburst seen in HD\,49330 (B0.5IVe) made using {\sl CoRot} data
\citep{Huat2009}. They found that the amplitudes of the $p$-modes
(short period) and $g$-modes (long period) were directly correlated
with the outburst: the amplitude of the $p$-modes decreased just
before the outburst and increased after the start of outburst. This is
almost identical to our findings for HD\,249179. Indeed, there is
evidence that the combination of pulsation modes actually cause
outbursts from Be stars (e.g. \citet{Baade2017}), with the increase in
the amplitude of the pulsations during the outburst has been explained
by stochastically excited waves
(e.g. \citet{Baade2018,Neiner2020}). More detailed work would be
needed to identify the modes in HD\,249179 before a similar study
could be undertaken on the {\tess} outburst data.

It is not clear whether HD 249179 is a binary system. Unfortunately,
there are no radial velocity data for it in Gaia DR3, presumably
since it is in the Galactic plane ($b=1.9^{\circ}$) where blending of
the RVS spectra can be a serious issue. Studies on the
  multiplicity of early type stars (see \citet{Sana2017} for an
  overview) suggest the multiplicity fraction of B4--9 type stars in
  the Milky Way is $\sim$1/5. It is therefore a matter of speculation
of whether the outburst seen in HD 249179 is due to tidal interaction
of a companion star.

\section{Conclusions}

We have shown that each of the 23 optical counterparts to the HMXBs in
this survey show clear evidence of quasi-periodic modulations with
periods in the range $\sim$0.1--1 d which fully supports the earlier
ground based observations of HMXBs by \citet{GutierrezSoto2011} and
\citet{Schmidtke2014,Schmidtke2016,Schmidtke2019}. Many other massive
stars, including Be stars, also show quasi-periodic pulsations with
similar periods. To be able to derive fundamental parameters of the
stars it is essential to identify the modes in the power
spectrum. This is not normally achievable just from broad band
photometry, but generally must be combined with high resolution
spectra (e.g. \citet{Aerts2019}). However, this will be much more
difficult to achieve for HMXBs where an accretion disc will make it
difficult to separate the disc and stellar pulsation
emission. Observations of massive eclipsing binaries
(e.g. \citet{Southworth2020}) show that tidal forces can affect the
pulsation frequencies. Given that some HMXBs, primarily those
  with Be donor stars, have a non-negligible eccentricity (e.g. Vela
X-1 has $e\sim$0.09) it is possible that a more detailed study of the
pulsation spectrum as a function of their orbital period may reveal
some dependency on the orbital phase. Finally, in Cycle 5, the cadence
of {\tess} full frame images will be 200 s which will allow many
more HMXBs and early type stars to be searched for pulsations.

\section{Acknowledgments}

This paper includes data collected by the {\sl TESS} mission, for
which funding is provided by the NASA Explorer Program.  This work
presents results from the European Space Agency (ESA) space mission
{\sl Gaia}. {\sl Gaia} data is being processed by the {\sl Gaia} Data
Processing and Analysis Consortium (DPAC). Funding for the DPAC is
provided by national institutions, in particular the institutions
participating in the {\sl Gaia} MultiLateral Agreement (MLA). The Gaia
mission website is \url{https://www.cosmos.esa.int/gaia}. The Gaia
archive website is \url{https://archives.esac.esa.int/gaia}.  This
research has made use of MAXI data provided by RIKEN, JAXA and the
MAXI team. Armagh Observatory \& Planetarium is core funded by the
Northern Ireland Executive through the Dept for Communities. We
  thank the anonymous referee for a detailed review of the
  manuscript.

\section*{Data Availablity}

The {\tess} data are available from the NASA MAST
portal\footnote{\url{https://archive.stsci.edu/}} and MAXI data 
  are available from the the Riken MAXI
portal.\footnote{\url{http://maxi.riken.jp}}

\appendix

\section{Light curves}

The light curves of all the HMXB in our sample where we indicate the
{\tess} sector the originated. For most cases we show the first sector
of data where more than one sector of data was obtained.

\begin{figure*}
    \centering
    \includegraphics[width=0.9\textwidth]{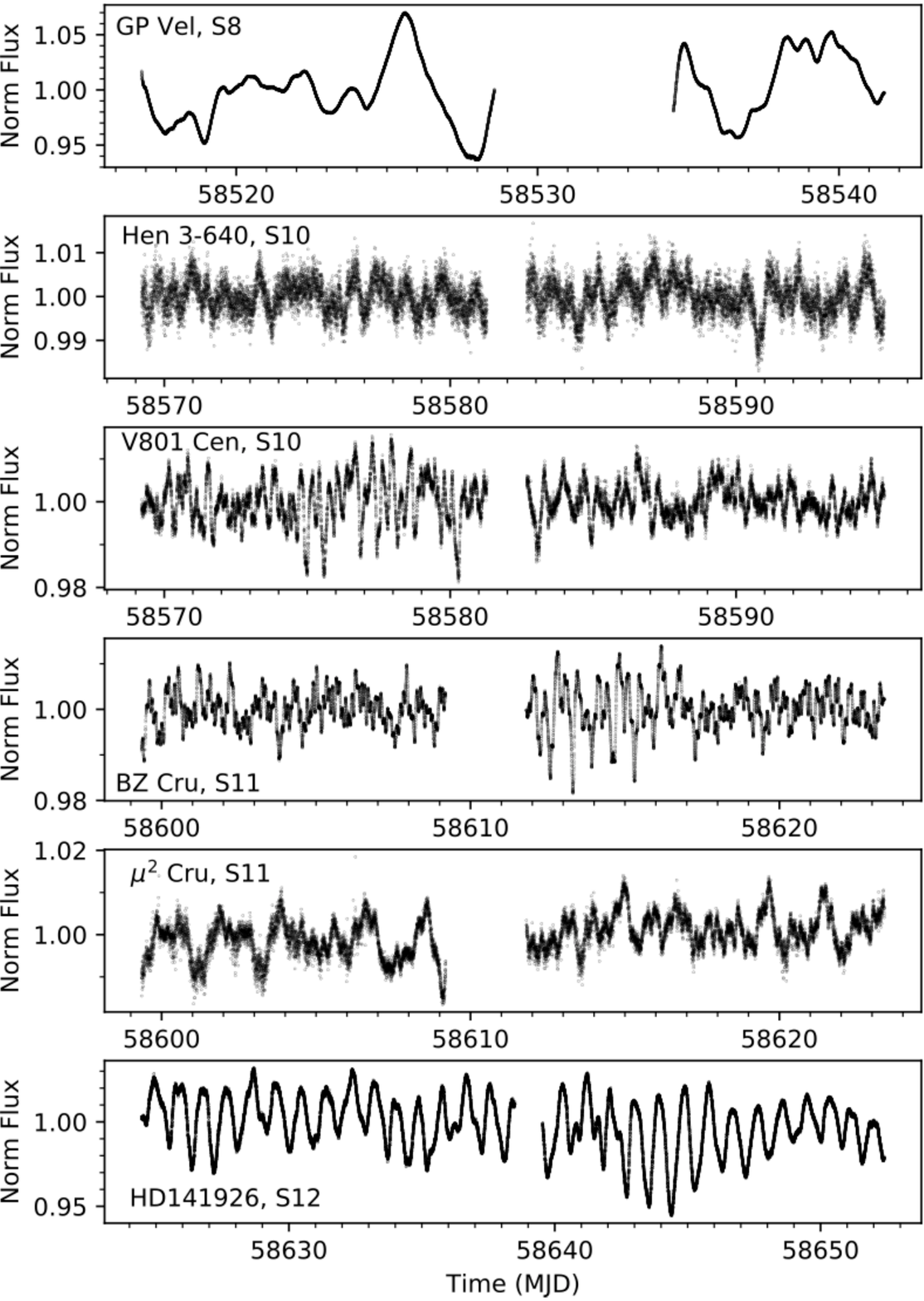}
    \caption{The light curves of the {\tess} light curves of the HMXBs
      in our sample where we indicate which sector the data originated.}
    \label{lightcurves-A}
\end{figure*}

\begin{figure*}
    \centering
    \includegraphics[width=0.9\textwidth]{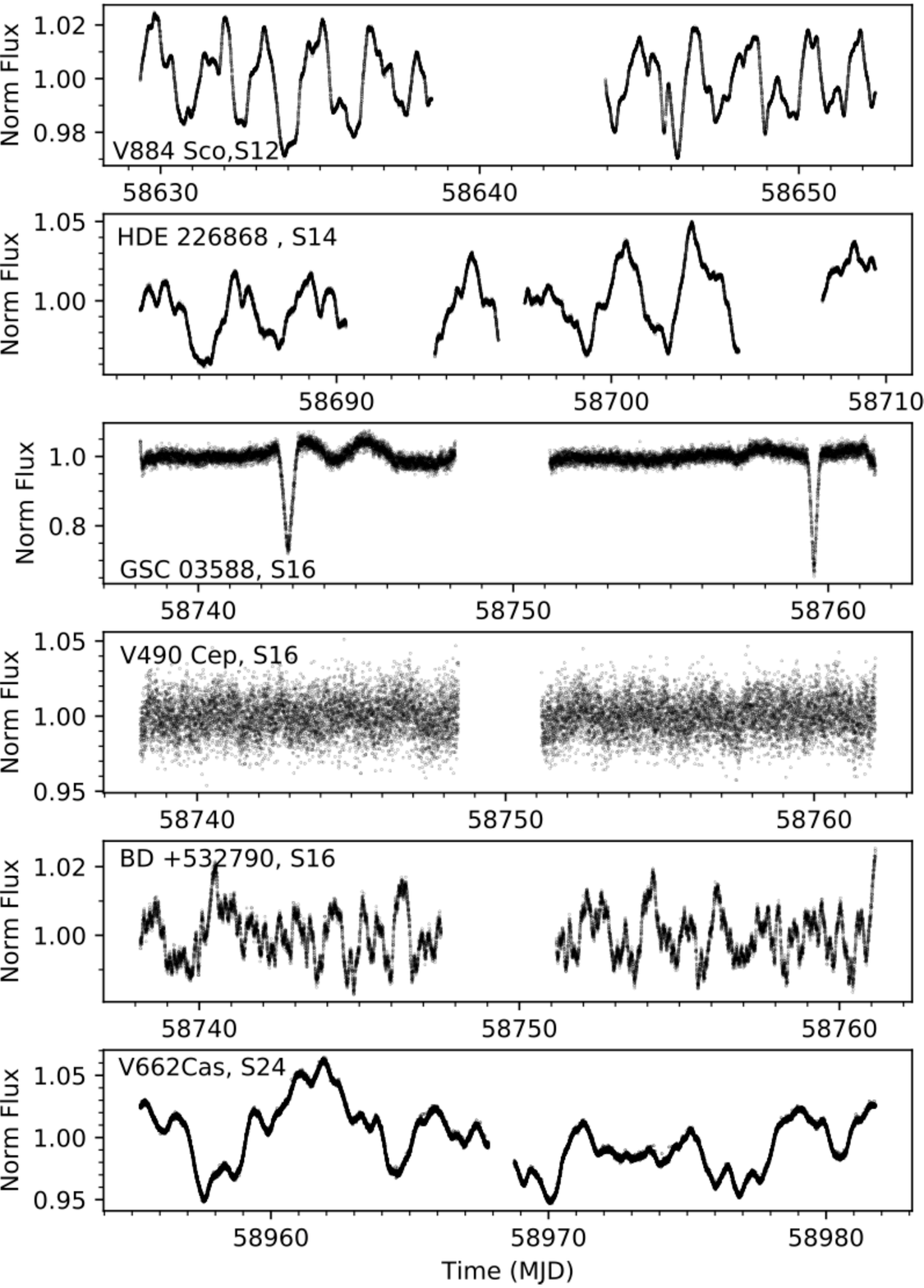}
    \caption{Figure \ref{lightcurves-A} continued.}
    \label{lightcurves-B}
\end{figure*}

\begin{figure*}
    \centering
    \includegraphics[width=0.9\textwidth]{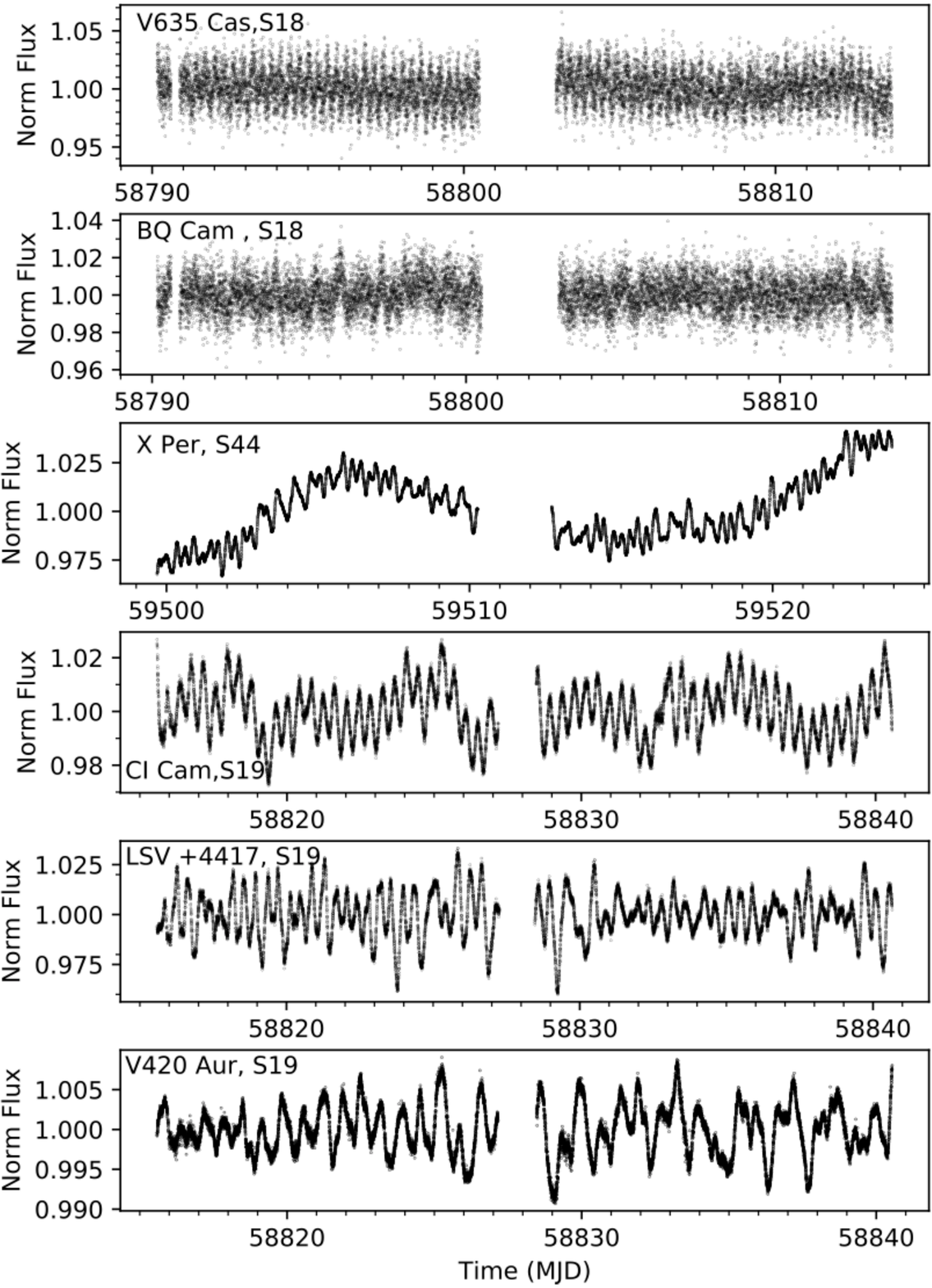}
    \caption{Figure \ref{lightcurves-B} continued.}
    \label{lightcurves-C}
\end{figure*}

\begin{figure*}
    \centering
    \includegraphics[width=0.9\textwidth]{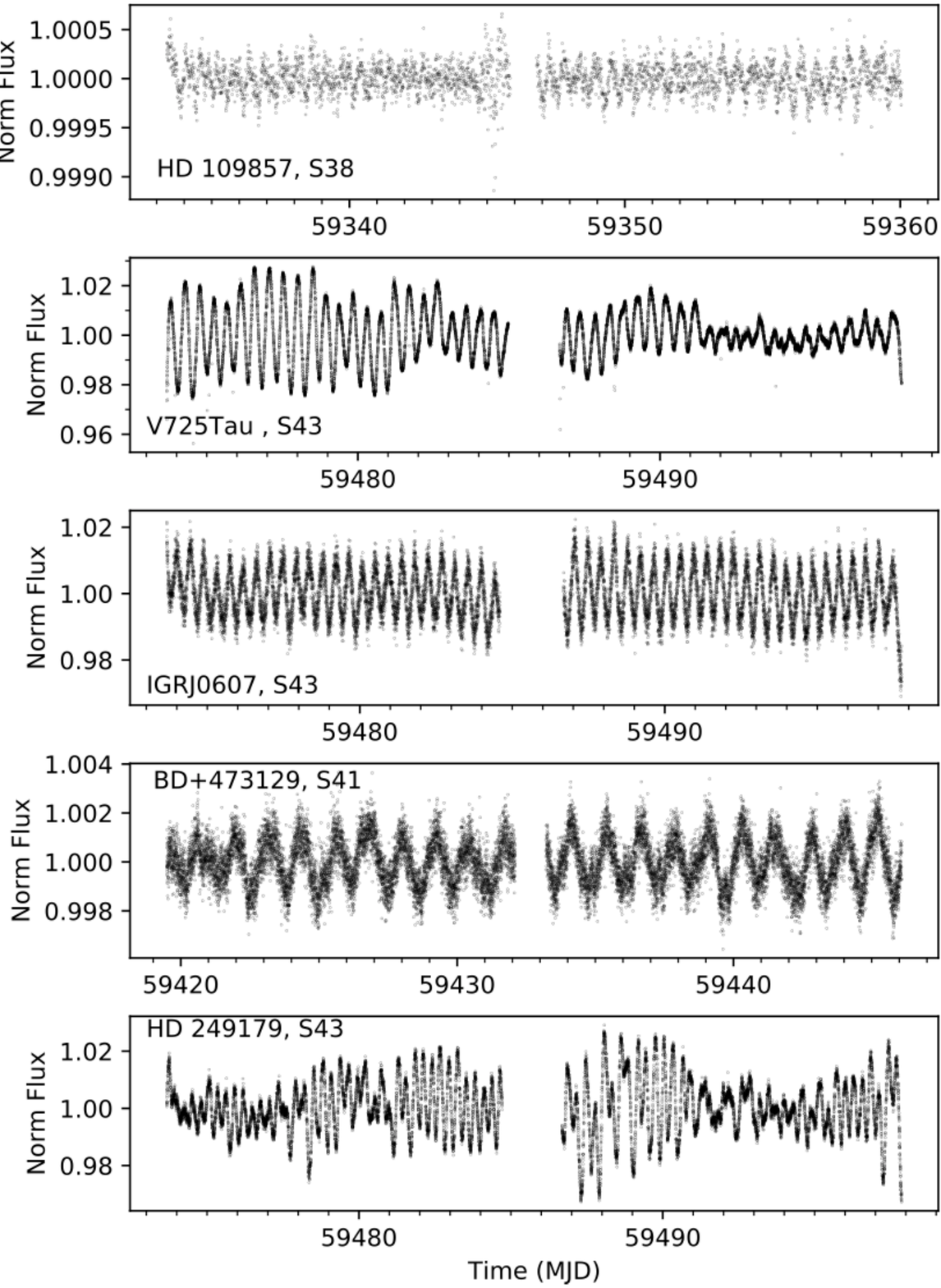}
    \caption{Figure \ref{lightcurves-C} continued.}
    \label{lightcurves-D}
\end{figure*}

\section{Detrended light curves}

For sources showing high amplitude variations in their light curve
over the orbital period we detrended the signature of the orbital
period to search for short period pulsations. In Figure
\ref{v635cas-cygx1} we show the original and detrended light curve of
HDE\,226868 (Cyg X-1). Here we show the original and detrended light
curves of GP\,Vel, V884\,Sco and V662\,Cas.

\begin{figure*}
%\vspace{0.5cm}
  \centering
    \includegraphics[width = 0.85\textwidth]{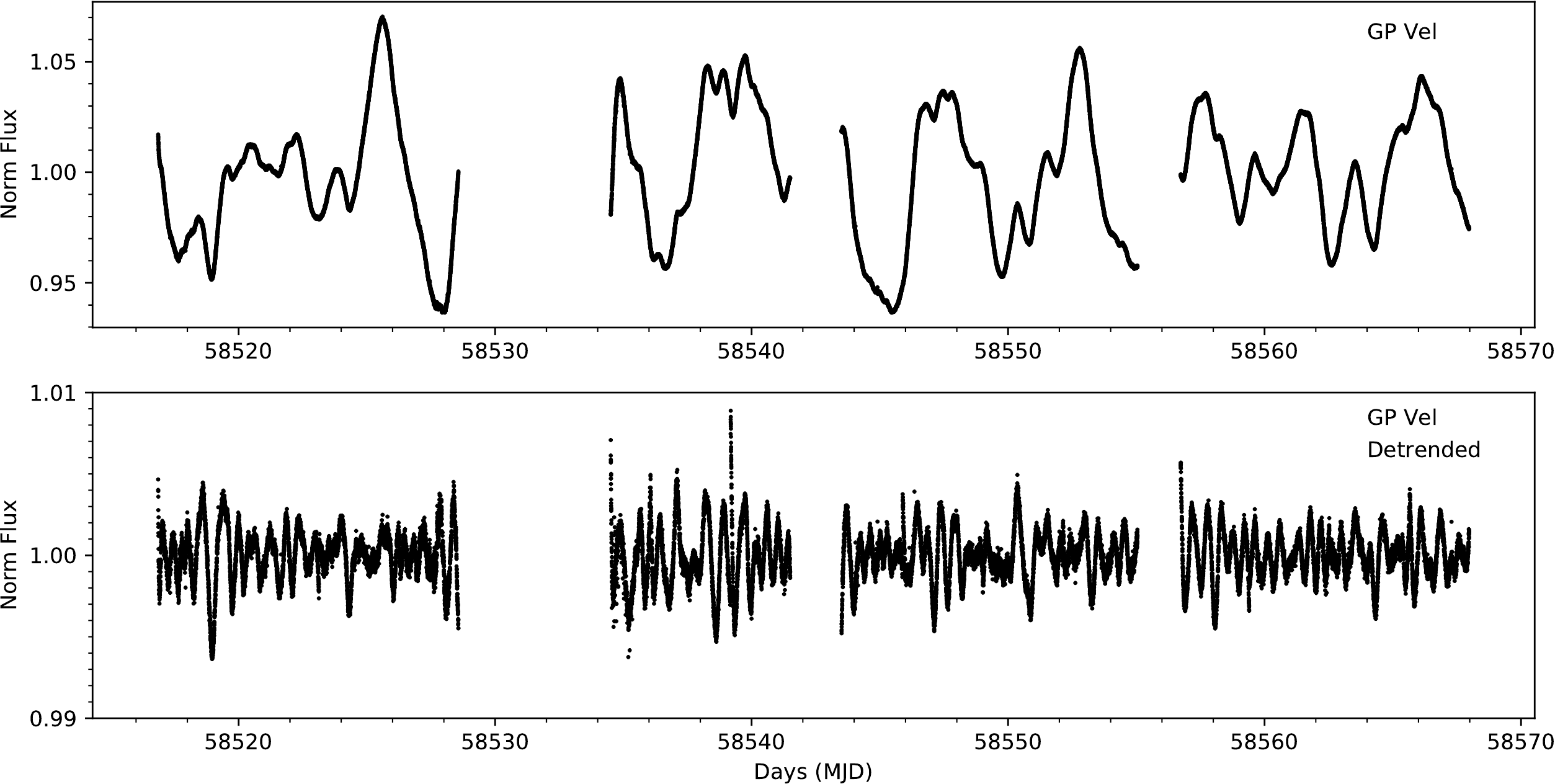}
    \includegraphics[width = 0.85\textwidth]{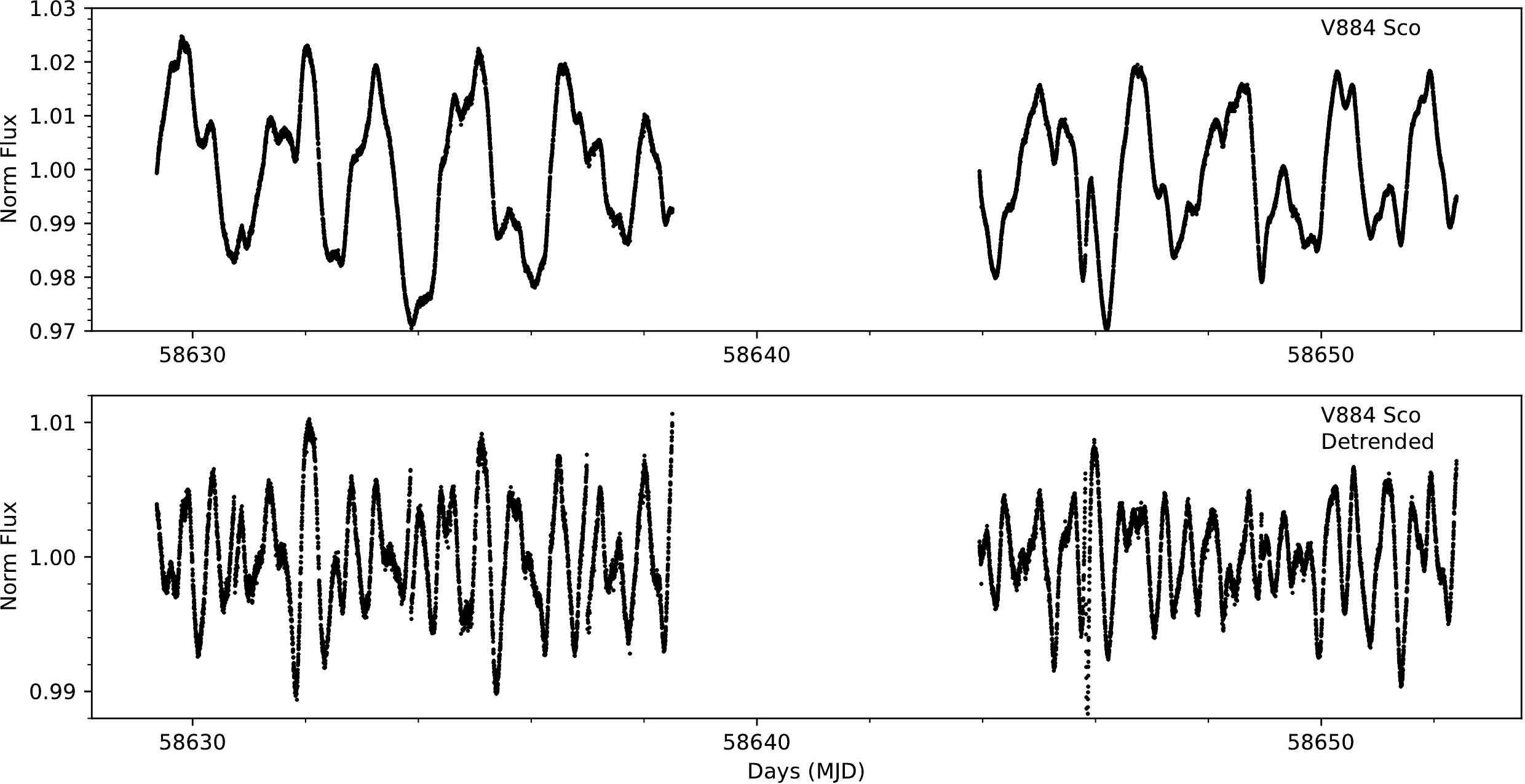}
    \includegraphics[width = 0.85\textwidth]{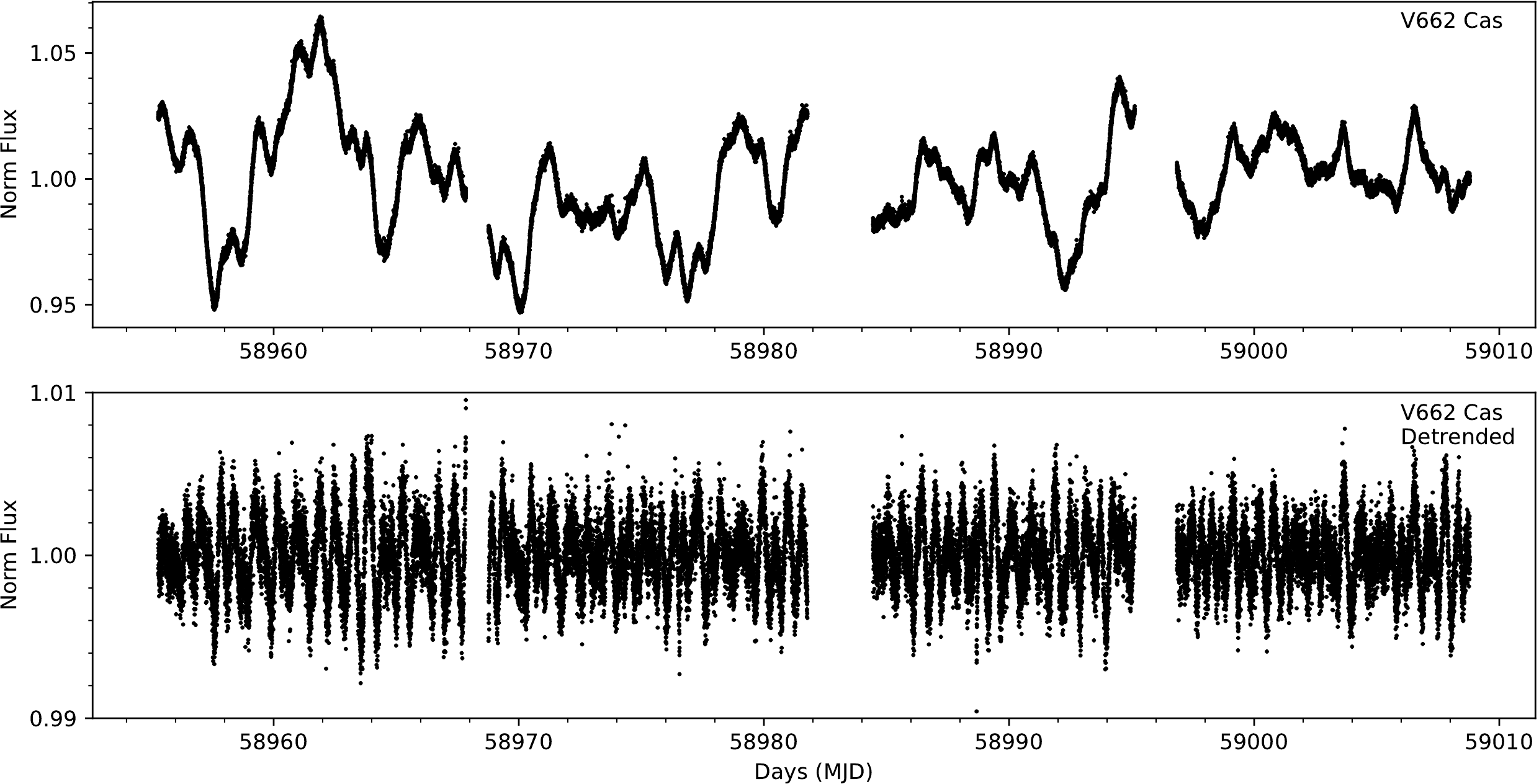}
    \caption{The original and detrended light curves of GP Vel; V884 Sco and V662 Cas.}
    \label{detrendappendix}
\end{figure*}

\section{Light Curves}

The light curves of V725 Tau and HD 249179 which have been detrended
to remove the signature of the outburst. The time of the optical
outbursts are indicated for each source.

\begin{figure*}
    \centering
    \includegraphics[width = 0.9\textwidth]{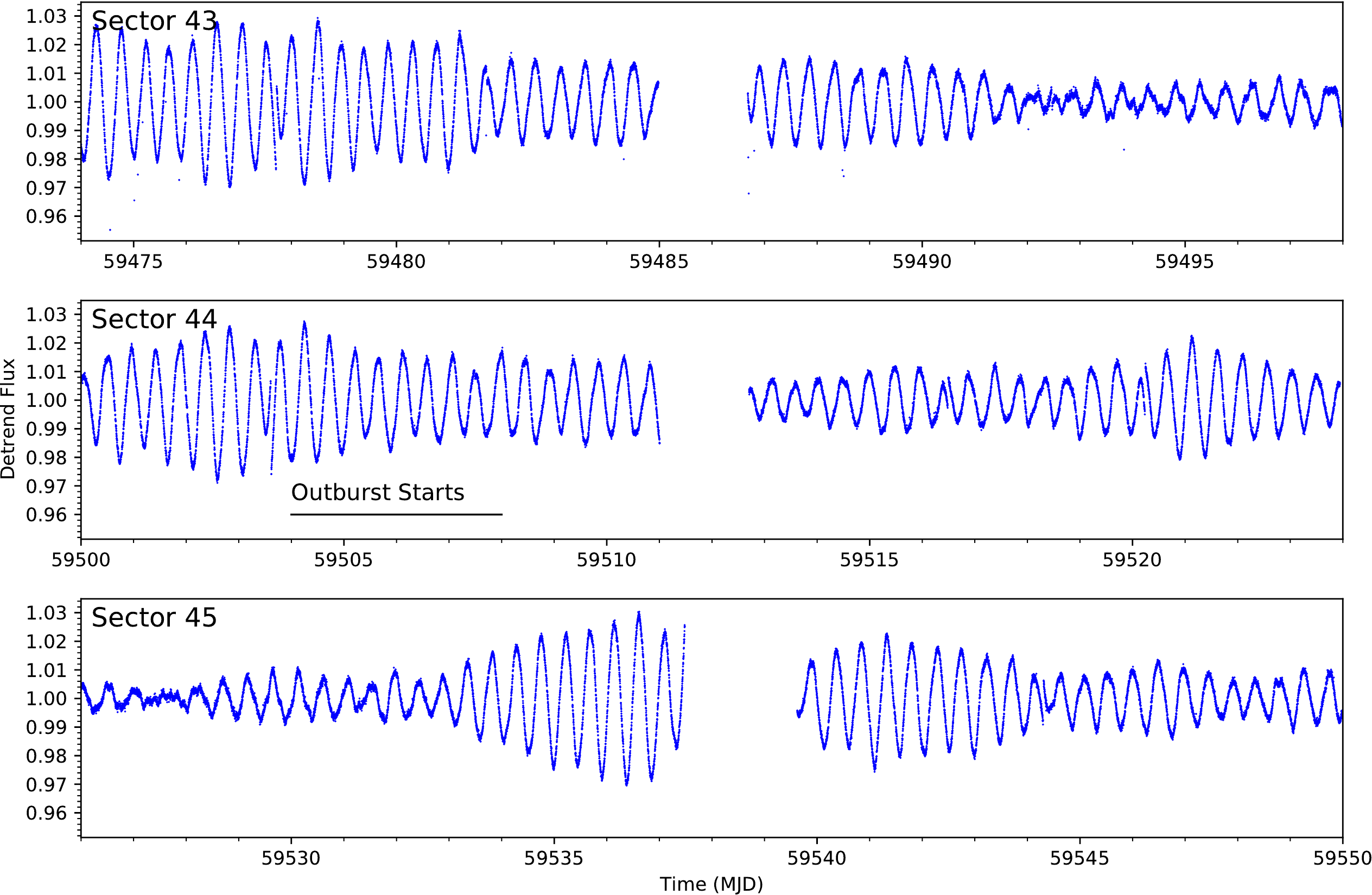}
    \caption{The {\tess} light curve of V725 Tau which has been
      detrended to remove the outburst and long-term
      variations. The resulting curve has a prominent signal at
        $\sim$0.48 d.}
    \label{V725-detrend}
\end{figure*}

\begin{figure*}
    \centering
    \includegraphics[width = 0.9\textwidth]{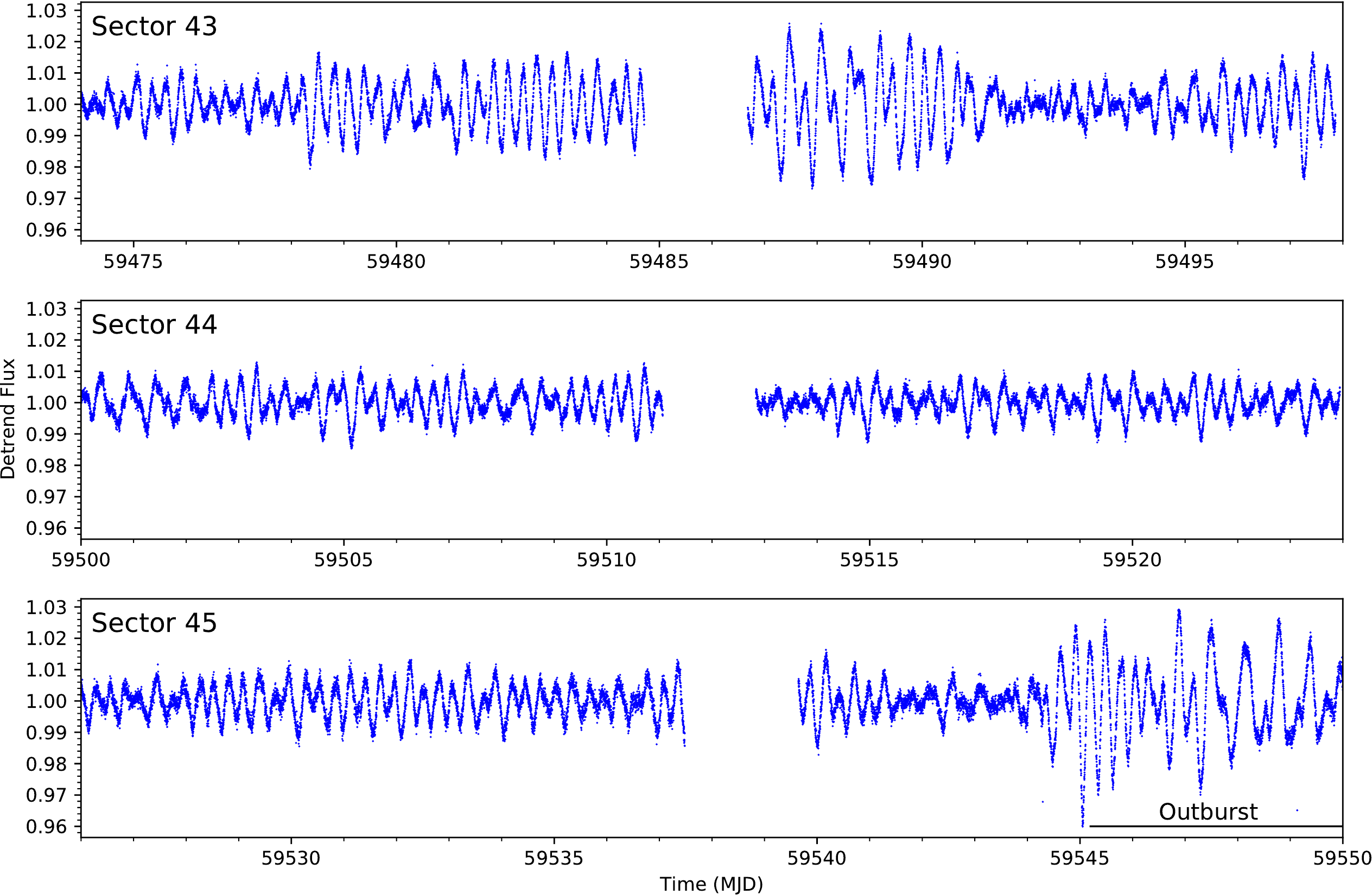}
    \caption{The {\tess} light curve of HD249179 which has been
      detrended to remove the outburst and long-term
      variations. The resulting curve has a prominent signal at
        $\sim$0.28 d}.
    \label{HD249179-detrend}
\end{figure*}

\end{document}